\documentclass[aps]{revtex4}
\usepackage{graphicx}
\begin{document}
\title{Rotating compact star with superconducting quark matter}
\author{P. K. Panda}
\author{H. S. Nataraj}
\affiliation{Indian Institute of Astrophysics, Koramangala,
Bangalore-560034, India}
\begin{abstract}
A compact star with superconducting quark core, the hadron crust and
the mixed phase between the two is considered. The quark meson
coupling model for hadron matter and the color flavor locked quark
model for quark matter is used in order to construct the equation of
state for the compact star. The effect of pairing of quarks in the
color flavor locked phase and the mixed phase on the mass, radius,
and period of the rotating star is studied.
\end{abstract}
\date{\today}
\maketitle
\vspace{0.2cm} PACS number(s): {95.30.Tg, 21.65.+f, 12.39.Ba,
25.75.Nq, 21.80.+a}

\vspace{0.2cm}
\section{Introduction}
Neutron/quark stars are the stellar objects produced as a result of
supernova explosions. These objects are highly compact as their
masses may, generally, be one or two times the solar mass and
radii of only about $10-15$ km. The neutron stars are normally
treated as zero temperature stellar objects because their
temperatures, although high, are still very low compared to their
characteristic energies of excitation. Their outer crust is normally
believed to be composed, mainly, of nuclei and electrons. 
Their interior, however,
where density is of the order of five to ten times the nuclear
saturation density, remains to be properly understood whether the
central part of the star is either composed of quark matter alone, 
or of mixed matter, or of paired quark matter is one of the 
subjects of the present work.

Recently, Alford and Rajagopal \cite{alford} discussed the difficulty 
of the 2-flavor superconducting (2SC) phase to achieve charge neutrality 
and therefore they claimed that this phase would not be realised in 
compact stars. Without solving the gap equation, they considered strange 
quark mass as a parameter in their calculations and concluded that the 
color flavor locked (CFL) phase would be favourable against 2SC. 
In Ref. \cite{steiner}, Steiner, Reddy and Prakash
investigated the color superconducting phase of quark matter by utlizing the 
Nambu-Jona-Lasino (NJL) model suplemented by diquark and the t'Hooft 
six fermion interactions. They found that in the NJL model, a small
2SC window does exist at very small baryon densities which eventually
would be shut by the hadronic phase. Again, the 2SC phase is less likely
to occur in compact stars because the charge neutrality condition
imposes a strong constraint to the quark chemical potential.
Further, in a recent work \cite{alford04} suggests that 
when the density drops low enough 
so that the mass of the strange quark can no longer be neglected, there 
is a continuous phase transition from the CFL phase to a new gapless CFL
(gCFL) phase, which could lead observable consequences if it occured
in the cores of neutron star \cite{alford05}. However, 
it now appears that some of the
gluons in the gCFL phase have imaginary Meissner masses indicating 
towards an unknown lower energy phase \cite{huang, casal}. 

The observations of the fastest pulsars with high degree of
compactness and the better theoretical understandings of the
matter at high densities like that of a compact star with
superconducting quark core in a color flavor locked phase
are compelling theoreticians to come up with
different models for the equation of state (EOS) of neutron stars
and their interiors.

In the present paper, we are interested in building the EOS for
mixed matter of quark and hadron phases. We employ the quark-meson
coupling model (QMC)~\cite{guichon88,ST} including hyperons
in order to describe the hadron phase. In the QMC model, baryons are
described as a system of non-overlapping MIT bags which interact
through the effective scalar and vector mean fields, very much in
the same way as in the Walecka model (QHD)~\cite{qhd}. Many
applications and extensions of the model have been made in the last few
years~\cite{recent}.

While the QMC model shares many similarities with QHD-type models,
it however offers new opportunities for studying nuclear matter
properties. One of the most attractive aspects of the model is that
different phases of hadronic matter from very low to very high
baryon densities and temperatures can be described within the same
underlying model. This model describes nuclear matter with nucleons
as nonoverlapping MIT bags and the quarks inside them couple to
scalar and vector mesons. For matter at very high density and/or
temperature, one expects that baryons and mesons dissolve and the
entire system of quarks and gluons becomes confined within a single,
big MIT bag. Another important aspect of the QMC is that the
internal structure of the nucleon is introduced explicitly. It is
found that the EOS for infinite nuclear matter at zero temperature
derived from the QMC model is much softer than the one obtained in
the Walecka model~\cite{qhd}. Also, the QMC model nucleon effective
mass lies in the range $ 0.7$ to $0.8$ of the free nucleon mass,
which agrees with results derived from non-relativistic analysis of
scattering of neutrons from lead nuclei~\cite{mahaux87} and is
larger in comparison with Walecka model effective mass.

For the quark phase, we consider pairing of quarks described by
the color-flavor locked (CFL) phase. Recently many authors
~\cite{cfl} have discussed the possibility
of quark matter in a color-superconducting phase, wherein
quarks near the Fermi surface are paired and these Cooper pairs
condense and break the color gauge symmetry~\cite{mga}. At
sufficiently high density the favored phase is called CFL, in which
quarks of all three colors and all three flavors are allowed to
pair.

We organize the paper as follows: In section 2, we briefly discuss i
the construction of the equation of
state for a compact star. The mixed EOS would be built by enforcing
appropriate Gibbs criteria and chemical equilibrium conditions. In
section 3, we present our numerical results and discuss the
properties like the maximum mass, radius and the period of the compact
star, etc. under the influence of the pairing gap, $\Delta$ and the
bag pressure, $B$.
\section{Equation of state}
We now discuss the EOS for a compact star with a minimal
mathematical details. The details of the QMC model can be found in
~\cite{guichon88,ST} and the details of CFL model can be
found in~\cite{cfl}.

In QMC model, the nucleon in nuclear medium is
assumed to be a static spherical MIT bag in which quarks interact
through the scalar and vector, $\sigma$, $\omega$ and $\rho$
mesons. And these fields are treated as classical fields in the mean
field approximation. The quark field, $\psi_q(x)$, inside the bag
then satisfies the equation of motion:
\begin{equation}
\left[i\,\rlap{/}\partial-(m_q^0-g_\sigma^q\, \sigma)-g_\omega^q\,
\omega\,\gamma^0 + \frac{1}{2} g^q_\rho \tau_z \rho_{03}\right]
\,\psi_q(x)=0\ , \quad  q=u,d,s, \label{eq-motion}
\end{equation}
where $m_q^0$ is the current quark mass and $g_\sigma^q$,
$g_\omega^q$ and $g_\rho^q$ are the quark-meson coupling
constants. After enforcing the boundary condition at the bag
surface, the transcendental equation for the ground state solution
of the quark (in $s$-state) is $ j_0(x_q)=\beta_q\, j_1(x_q)\;$,
which determines the bag eigenfrequency $x_q$, where
$\beta_q=\sqrt{(\Omega_q-R_B m_q^*)/(\Omega_q+R_B m_q^*)}$, with
$\Omega_q=(x^2_q+R^2{m^*_q}^2)^{1/2}$;
$m^*_q=m^0_q-g_\sigma^q\sigma_0$, is the effective quark mass.
The energy of a static bag describing baryon $B$ consisting of three
ground state quarks can be expressed as
\begin{equation}
M^*_B=\sum_q n_q \, {\Omega_q\over R_B}-{Z_B\over R_B} +{4\over 3}\,
\pi \, R_B^3\,  B_B\ , \label{ebag}
\end{equation}
where $B_B$ is the bag constant and $Z_B$ parameterizes the sum of
the center-of-mass motion and the gluonic corrections. The medium
dependent bag radius, $R_B$ is then obtained through the stability
condition for the bag.

The total energy density, $\varepsilon$, and the pressure, $p$,
including the leptons can be obtained from the grand canonical
potential and they read
\begin{eqnarray}
\varepsilon &=& \frac{1}{2}m_\sigma^2 \sigma^2
+ \frac{1}{2}m_\omega^2 \omega^2_0
+ \frac{1}{2} m_\rho^2 \rho^2_{03} \nonumber\\
&+& \sum_B \frac{\gamma}{2\pi^2} \int_0^{k_B}k^2 dk
\left[k^2 + M_B^{* 2}(\sigma)\right]^{1/2}
+ \sum_l \frac{1}{\pi^2} \int_0^{k_l} k^2  dk
\left[k^2 + m_l^2\right]^{1/2}~,
\end{eqnarray}
\begin{eqnarray}
p &=& - \frac{1}{2}m_\sigma^2 \sigma^2
+ \frac{1}{2}m_\omega^2 \omega^2_0
+ \frac{1}{2} m_\rho^2 \rho^2_{03} \nonumber\\
&+& \frac{1}{3} \sum_B \frac{\gamma}{2\pi^2} \int_0^{k_B}
\frac{k^4 \ dk}{\left[k^2 + M_B^{* 2}(\sigma)\right]^{1/2}}
+ \frac{1}{3} \sum_l \frac{1}{\pi^2} \int_0^{k_l} \frac{k^4  dk}
{\left[k^2 + m_l^2\right]^{1/2}} ~.
\end{eqnarray}
In the above equations $\gamma$ and $k_B$ are respectively
the spin degeneracy and the Fermi momentum of the baryon
species $B$. The hyperon couplings, $g_{\sigma B}=x_{\sigma B}~
g_{\sigma N},~~g_{\omega B}=x_{\omega B}~ g_{\omega N}, ~~g_{\rho
B}=x_{\rho B}~ g_{\rho N}$ are not relevant to the ground state
properties of nuclear matter, but information about them can be
obtained from the levels in $\Lambda$ hypernuclei~\cite{chrien89}.
The $x_{\sigma B}$, $x_{\omega B}$ and $x_{\rho B}$ are equal to $1$
for nucleons and acquire different values in different
parameterizations for the other baryons. Note that the $s$-quark is
unaffected by the $\sigma$ and $\omega-$ mesons i.e.
$g_\sigma^s=g_\omega^s=0\ .$ The lepton Fermi momenta are the
positive real solutions of $(k_e^2 + m_e^2)^{1/2} =  \mu_e$ and
$(k_\mu^2 + m_\mu^2)^{1/2} = \mu_\mu = \mu_e$. The equilibrium
composition of the star is obtained with the charge neutrality
condition at a given total baryonic density $\rho = \sum_B
\gamma~ k_B^3/(6\pi^2)$; the baryon effective masses are obtained
self-consistently in the bag model. For stars in which the strongly
interacting particles are baryons, the composition is determined by
the requirements of charge neutrality and $\beta$-equilibrium
conditions under the weak processes $B_1 \to B_2 + l + {\overline
\nu}_l$ and $B_2 + l \to B_1 + \nu_l$. After deleptonization, the
charge neutrality condition yields $ q_{\rm tot} = \sum_B q_B~
\gamma~ k_B^3 \big/ (6\pi^2) + \sum_{l=e,\mu} q_l k_l^3 \big/ (3\pi^2)
= 0 ~, $ where $q_B$ corresponds to the electric charge of baryon
species $B$ and $q_l$ corresponds to the electric charge of lepton
species $l$. Since the time scale of a star is effectively infinite
compared to the weak interaction time scale, weak interaction
violates strangeness conservation. The strangeness quantum number is
therefore not conserved in a star and the net strangeness is
determined by the condition of $\beta$-equilibrium which for baryon
$B$ is, then, given by $\mu_B = b_B\mu_n - q_B\mu_e$, where $\mu_B$ is
the chemical potential of baryon $B$ and $b_B$ is its baryon number.
Thus the chemical potential of any baryon can be obtained from the
two independent chemical potentials $\mu_n$ and $\mu_e$ of neutron
and electron respectively.

We next study the equation of state of a compact star taking into
consideration the CFL quark phase. We treat the quark matter as a
Fermi sea of free quarks with an additional contribution to the
pressure arising from the formation of CFL condensates.

The CFL phase can be described with the help of the thermodynamical
potential which reads~\cite{ar}:
\begin{equation}
\Omega_{CFL}(\mu_q,\mu_e)=\Omega_{quark}(\mu_q) +
\Omega_{GB}(\mu_q,\mu_e) + \Omega_{l}(\mu_e),
\end{equation}
where $\mu_q=\mu_n/3$ and
\begin{equation}
\Omega_{quark}(\mu_q)=\frac{6}{\pi^2}
\int_0^{\nu} p^2 d p (p-\mu_q) +\frac{3}{\pi^2}
\int_0^{\nu} p^2 d p (\sqrt{p^2 + m_s^2}-\mu_q)
- \frac{3 \Delta^2 \mu_q^2}{\pi^2} +B,
\end{equation}
with  $m_u=m_d$ set to zero,
\begin{equation}
\nu=2 \mu_q - \sqrt{\mu_q^2 + \frac{m_s^2}{3}},
\end{equation}
$\Omega_{GB}(\mu_q,\mu_e)$ is the contribution from the Goldstone
bosons arising due to the chiral symmetry breaking in the CFL phase
~\cite{ar,son}:
\begin{equation}
\Omega_{GB}(\mu_q,\mu_e)=-\frac{1}{2} f_{\pi}^2 \mu_e^2 \left(1 -
\frac{m_\pi^2}{\mu_e^2} \right)^2, \label{ogb}
\end{equation}
where
\begin{equation}
f_{\pi}^2=\frac{(21-8~~ ln 2)\mu_q^2}{36 \pi^2}
~~~,~~~
m_\pi^2= \frac{3 \Delta^2}{\pi^2 f_\pi^2} m_s (m_u + m_d),
\end{equation}
$\Omega_{l}(\mu_e)=-{(\mu_e)^4}/{12 \pi^2}$, and the quark
number densities are equal, i.e.,
\begin{equation}
\rho_u=\rho_d=\rho_s=\frac{\nu^3 + 2 \Delta^2 \mu_q}{\pi^2}.
\end{equation}
 and in the above expressions $\Delta$ is the gap parameter \cite{ar}.

We next consider the scenario of a mixed phase of hadronic and quark
matter where the hadron phase is described by the QMC model and the
quark phase is described by the CFL model. In the mixed phase charge
neutrality is imposed globally i.e. the quark and hadron phases
are not neutral separately but rather, the system prefers to
rearrange itself so that
\begin{equation}
\chi\rho_c^{QP}+(1-\chi)\rho_c^{HP}+\rho_c^{l}=0
\end{equation}
where $\rho_c^{QP}$ and $\rho_c^{HP}$ are the charge densities of
quark and hadron phases, respectively. $\chi$ is the volume fraction
occupied by the quark phase, $(1-\chi)$ is the volume fraction
occupied by the hadron phase and $\rho_c^{l}$ is the lepton charge
density. As usual, the phase boundary of the coexistence region
between the hadron and quark phase is determined by the Gibbs
criteria. The critical neutron and electron chemical potentials and
the critical pressure are determined by the conditions,
$\mu^{HP}_i=\mu^{QP}_i=\mu_i, \,\, i=n,e, \quad T^{HP}=T^{QP}, \quad
P^{HP}(\mu^{HP},T)=P^{QP} (\mu^{QP},T)$ reflecting the need of
chemical, thermal and mechanical equilibrium, respectively. The
energy density and the total baryon density in the mixed phase read:
\begin{equation}
\varepsilon=\chi\varepsilon^{QP}+(1-\chi)\varepsilon^{HP}+\varepsilon^{l},
\quad\quad \rho=\chi\rho^{QP}+(1-\chi)\rho^{HP}.
\end{equation}

\section{Results and discussion}

We start by fixing the free-space bag properties for the QMC model.
In the present calculation, we have used the current quark masses,
$m_u=m_d=0$ MeV and $m_s = 150$ MeV and $R_0=0.6$ fm for the bag
radius. There are two unknowns, $Z_B$ and the bag constant $B_B$.
These are obtained as usual by fitting the nucleon mass, $M=939$ MeV
and enforcing the stability condition for the bag. The values
obtained for $Z_B$ and $B_B$ are given in~\cite{panda04}. Next we
fit the quark-meson coupling constants $g_\sigma = g_\sigma^q$, $g_\omega =
3g_\omega^q$ and $g_\rho = g_\rho^q$ for the nucleon to obtain the
correct saturation properties of the nuclear matter, $E_B \equiv
\varepsilon/\rho - M = -15.75 \;$ MeV at $\rho = \rho_0=0.15$
fm$^{-3}$, $a_{sym}=32.5$ MeV, $K=257$ MeV and $M^*=0.774 M$. We
have taken $g_\sigma^q=5.957$, $g_{\omega N}=8.981$ and $g_{\rho
N}=8.651$. We take the standard values for the meson masses,
$m_\sigma=550$ MeV, $m_\omega=783$ MeV and $m_\rho=770$ MeV. The
meson-hyperon coupling constants used in our calculations are
$x_{\sigma\,B}=x_{\omega\,B}=x_{\rho\,B}=\sqrt{2/3}$ and are
obtained based on quark counting arguments~\cite{moszk}.

We first plot the pressure as a function of energy density in Figure
1 for different values of gap parameter, $\Delta$, keeping the bag
pressure, $B^{1/4}= 210$ MeV constant. We see that for lower the
pairing gap the higher would be the energy density and the pressure
of both the transition points between the hadron crust and quark
core. And the EOS of mixed phase is wider for lower $\Delta$. Also
the EOS for lower $\Delta$ in the mixed region is stiffer but it
becomes softer in the CFL region.

We next proceed to calculate the properties of the compact star
using the above EOS. The star is assumed to be rapidly rotating,
relativistic and compact. The details of the model are given by
Kamatsu {\it et. al}~\cite{keh}~[known as Komatsu, Eriguchi-Hachisu
(KEH) method] and Cook {\it et. al}~\cite{cook}. The star is assumed
to be stationarily rotating and hence have axially, equatorially
symmetric structures. The metric in spherical coordinates
$(t,r,\theta,\phi)$ can be written as,
\begin{equation}
ds^2=-e^{2\nu}dt^2+e^{2\alpha}(dr^2+r^2d\theta^2)+e^{2\beta}r^2
\sin^2\theta(d\phi-\omega dt)^2
\end{equation}
where the metric potentials, $\alpha$, $\beta$, $\nu$ and $\omega$
are functions of $r$ and $\theta$ only (geometrized units, $c=G=1$).
The matter is assumed to be a perfect fluid so that the
energy-momentum tensor $T^{ab}$ is given by
\begin{equation}
T^{ab}=(\varepsilon+p)U^aU^b+pg^{ab},
\end{equation}
where $\varepsilon$, $p$, $Uâ$, and $g^{ab}$ are the energy density,
pressure, four velocity and metric tensor respectively. It is
further assumed that the four velocity $U^a$ is simply a linear
combination of time and angular Killing vectors. The details of the
calculations for solving the field equations are given in Ref.
\cite{keh}.

The radius of the maximum mass star is sensitive to the low density
EOS. We have used Baym, Pethick and Sutherland~\cite{bps} values for
pressure and energy at low baryonic densities. In Figure 2, we have
shown the mass - radius relation for the hybrid star for different
values of $\Delta$ in the range $0-150$ MeV while keeping the bag
pressure constant, $B^{1/4}=~210$ MeV fixed. The maximum mass, $M_{max}$,
is higher for lower $\Delta$ and it shifts towards the low radius
side as $\Delta$ decreases. The $M_{max}$ for $\Delta = 150$ MeV is
$1.61~ M_{\odot}$ and the corresponding radius, $R_{max}$ is $18.54$
km, where as, for $\Delta = 0$ MeV $M_{max}$ and $R_{max}$ are
$2.35~ M_{\odot}$ and $16.47$ km respectively. For an unpaired
quark matter star \cite{uqm}, for the same bag pressure of 210 MeV, we
obtained maximum mass and radius, are $2.36~ M_{\odot}$ and $16.12$
km respectively. To compare with, van Kerkwijk et al. have obtained
the mass of $1.86\pm0.32$ for an X-ray pulsar Vela X-1~\cite{bkk}.
The compactness of a star, which is the ratio of $M_{max}$ to
$R_{max}$, therefore, decreases with increasing $\Delta$. We observe
from Figure 3 that, for $\Delta = 0$ MeV the ratio turns out to be
$0.142~ M_{\odot}/$km, where as, for $\Delta = ~150$ MeV it is
$0.086~ M_{\odot}/$km. The general relativistic limit for the ratio
is, \(M/R < 4/9\)~\cite{nkg} (i.e. for a uniform density star with
the causal equation of state $P=\varepsilon$) which is much above
our range. The lower limit for the ratio found from observations
of~\cite{sanwal} is $0.115 M_{\odot}/$km. Hence the hybrid star
prefers to have low pairing gap to be compact.

We observe that, as the bag pressure increases, keeping the pairing
gap fixed, both the central energy density, $\varepsilon_{c}$ and
$M_{max}$ increases but the radius, $R_{max}$ of a star decreases.
This is in consistent with~\cite{igor03}. For a fixed $\Delta$ of
$100$ MeV, $\varepsilon_{c} = 7.95 \times 10^{14}$ gcm$^{-3}$,
$M_{max} = 1.32\;M_{\odot}$, and $R_{max} =~ 18.8$ km for $B^{1/4} =
190$ MeV, where as, for $B^{1/4} = 210$ MeV, $\varepsilon_{c} = 1.29
\times 10^{15}$ gcm$^{-3}$, $M_{max}$ =~ 2.17$M_{\odot}$, and
$R_{max} = 17.4$ km. But when the bag pressure is kept constant, the
decrease in $\Delta$ increases $\varepsilon_{c}$. Our results thus
reflect that the dense cores of the compact stars favor low pairing
gap and high bag pressures.

We have also observed that, as $\Delta$ increases the $M_{max}$
decreases unlike the plots in~\cite{lug03} where they have shown an
opposite trend i.e, the increase in maximum mass with the pairing gap
for pure CFL stars with a broader $\Delta$ range they have considered.
We would like to mention that, our results are not contradicting
theirs inasmuch as we also have checked for the pure CFL phase and
have seen the same trend. From our results, we observe that the
stability of the hybrid star will not favor the onset of pure CFL
phase at its core, with the EOS we have used as the $M_{max}$ lies
in the mixed region as in~\cite{panda04} for all the values of
$\Delta$ in the range from $0 - 150$ MeV for a fixed bag pressure of
$210$ MeV. Thus, the CFL effect is reflected on to the stability of
the hybrid star only through the mixed phase.

In Figure 4, we have plotted the angular velocity versus the mass of
the  rotating compact star. For a fixed value of $\Delta$ and the
bag pressure of $B^{1/4}=~210$ MeV, the angular velocity increases
with the mass. And we see that as $\Delta$ decreases both the
angular velocity and the $M_{max}$ increases. For $\Delta = 150$
MeV, the angular velocity at $M_{max}$ is $5.5\times 10^3~ s^{-1}$,
where as, for $\Delta = 0$ MeV it is $7.7\times 10^3~ s^{-1}$. For an
unpaired quark matter star also the angular velocity obtained is
$7.7\times10^3~ s^{-1}$. The angular velocities for the fastest
observed pulsars till date, PSR $B1937+21$ and PSR $B1957+20$, are
$4033~ s^{-1}$ and $3908 s^{-1}$ respectively~\cite{thor}. We have
also seen that the angular velocity increases  with a decrease in
the radius of the compact star. The effect of the paring gap on the
period, $P = 2 \pi/\Omega$, of the rotating star, for a fixed bag
pressure of $B^{1/4}=~210$ MeV, is shown in Figure 5. As the pairing
gap increases the period increases. The limit imposed by general
relativity on the period of a relativistic compact star should be,
\(P > 0.24\) ms~\cite{nkg}. The period for the said fastest pulsars
till date is $1.55$ ms and $1.6$ ms respectively~
\cite{friedman95,thor}. The period in our case lies in the range of
$0.8$ to $1.2$ milliseconds for the $\Delta$ range of $0-150$ MeV.
The periods predicted for strange stars are in submillisecond
range~\cite{nkg}. Thus, the stars with low pairing energies are more
compact and are the fast rotors.

Let us now summarize our results. We have studied the compact stars with
a mixed matter of hadrons and deconfined quarks using QMC model for
hadron phase and CFL quark model for quark phase. We conclude here
that the dense hybrid stars indeed favor low pairing energy and high
bag pressures. And these stars are more massive, more compact, and
are fast rotors with the periods of sub-milliseconds. Also our
results clearly show that the hybrid stars, of the kind we have
considered, possess a quark matter core only as a mixed phase of
deconfined quark matter in a CFL phase and hadrons. And no pure CFL
quark matter core can be possible on the stability grounds. These
findings have to be deduced more carefully both from the
astrophysical observations of compact stars and from laboratory
measurements of high density matter to confirm or discard the ranges
of the said parameters we have considered.

\acknowledgments The authors acknowledge Professor B. P. Das,
Professor C. Provid\^encia and Professor D. P. Menezes for their
critical suggestions. PKP would like to thank the friendly
atmosphere at Indian Institute of Astrophysics, Bangalore, where
this work was partially done.

\newpage
\begin{figure}
\includegraphics[width=10.cm]{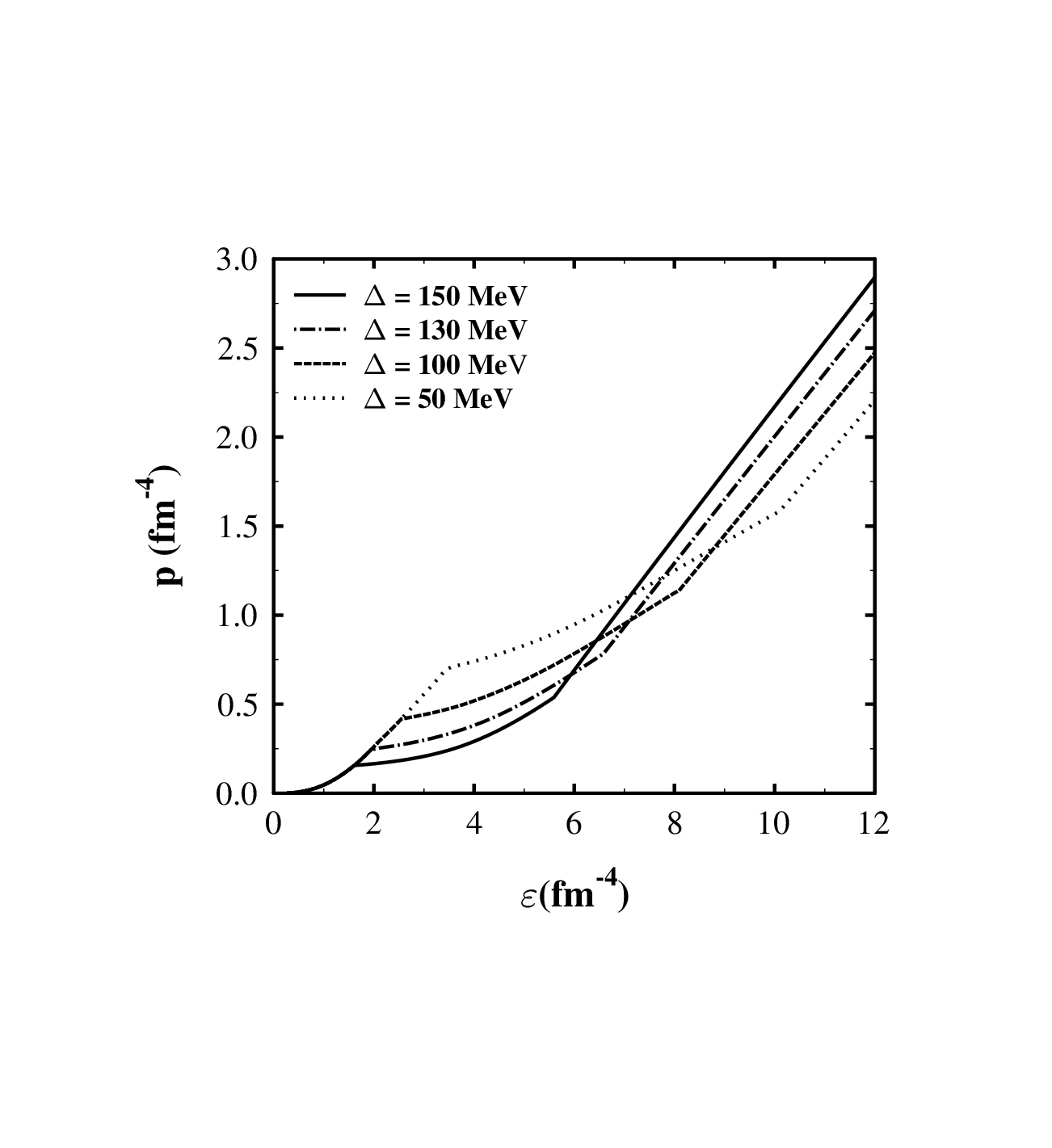}
\caption{Equation of state for different values of
pairing gap and for a fixed bag pressure, $B^{1/4}= 210$ MeV. The
stiffer(softer) EOS in mixed phase becomes softer(stiffer) in CFL
phase.} \label{eoscomp}
\end{figure}
\begin{figure}
\includegraphics[width=10.cm]{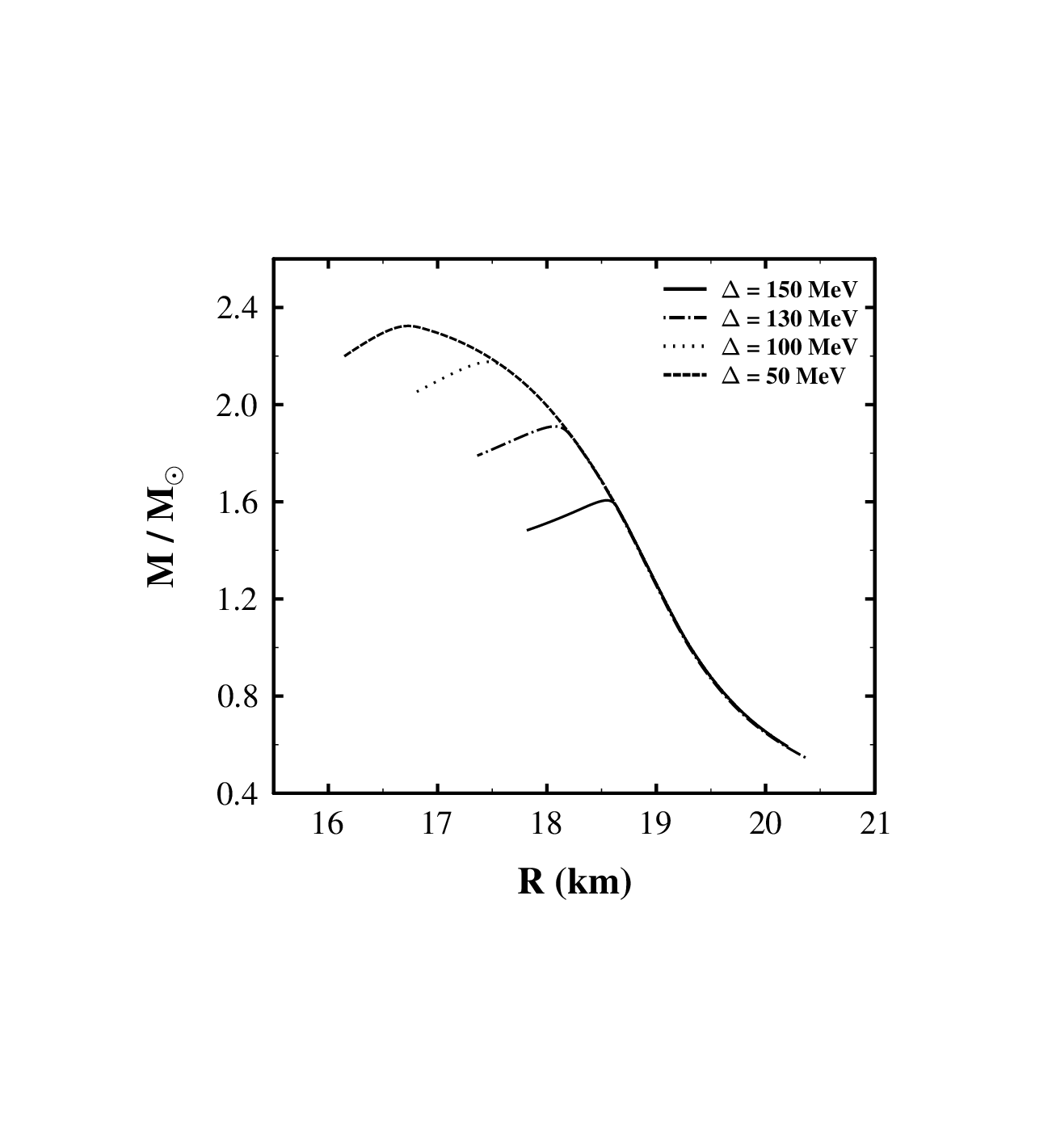}
\caption{The mass - radius relation for different
values of pairing gap and for a fixed bag pressure, $B^{1/4}= 210$
MeV. As the pairing gap decreases the $M_{max}$ increases and it
shifts towards lower radius side.} \label{mrcomp}
\end{figure}
\begin{figure}
\includegraphics[width=10cm]{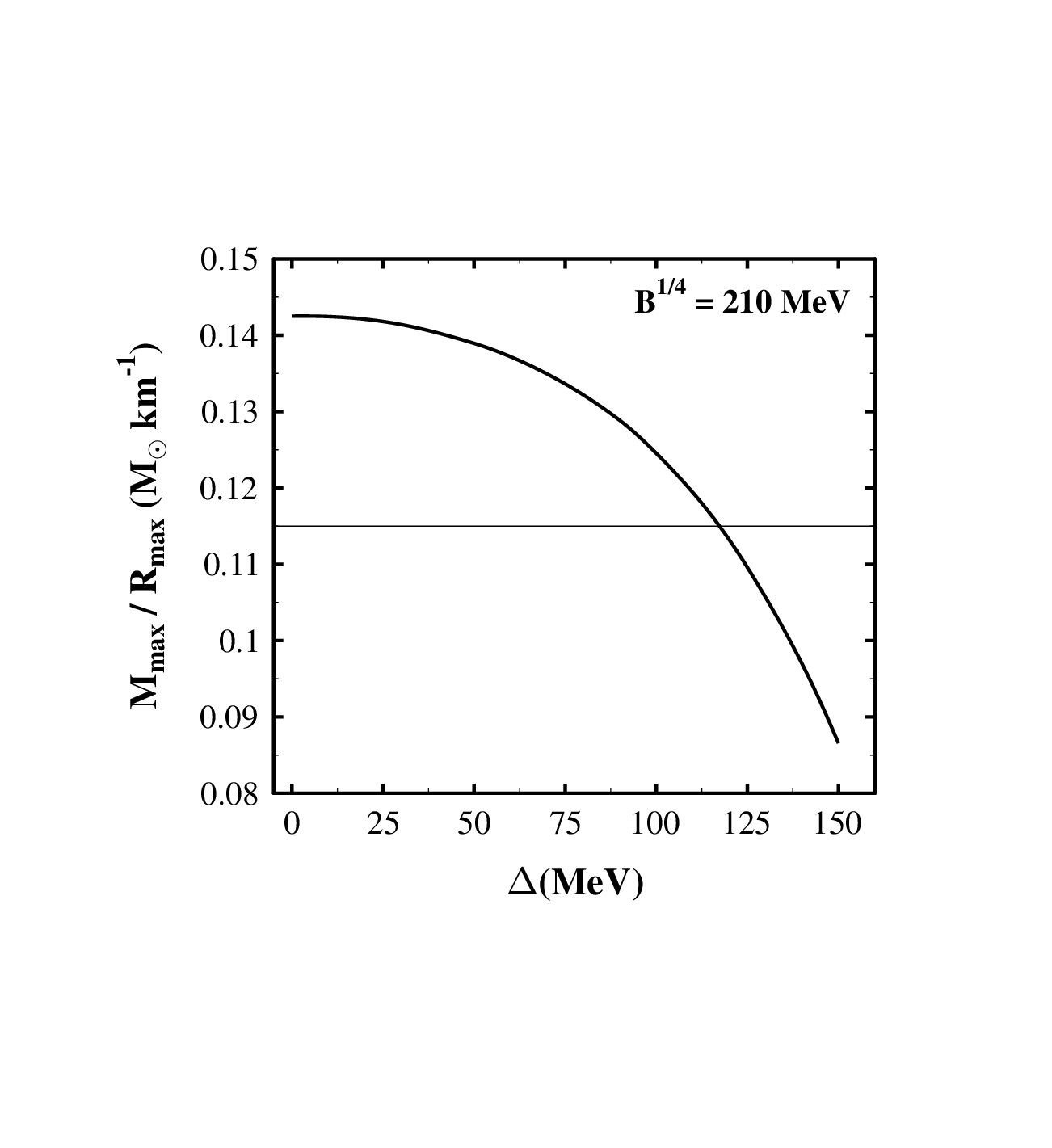}
\caption{The degree of compactness, which is the ratio of maximum
mass to the corresponding radius of the compact star, decreases with
an increasing pairing gap while keeping bag pressure, $B^{1/4}= 210$
MeV, fixed.} \label{mrdelta}
\end{figure}
\begin{figure}
\includegraphics[width=10.cm]{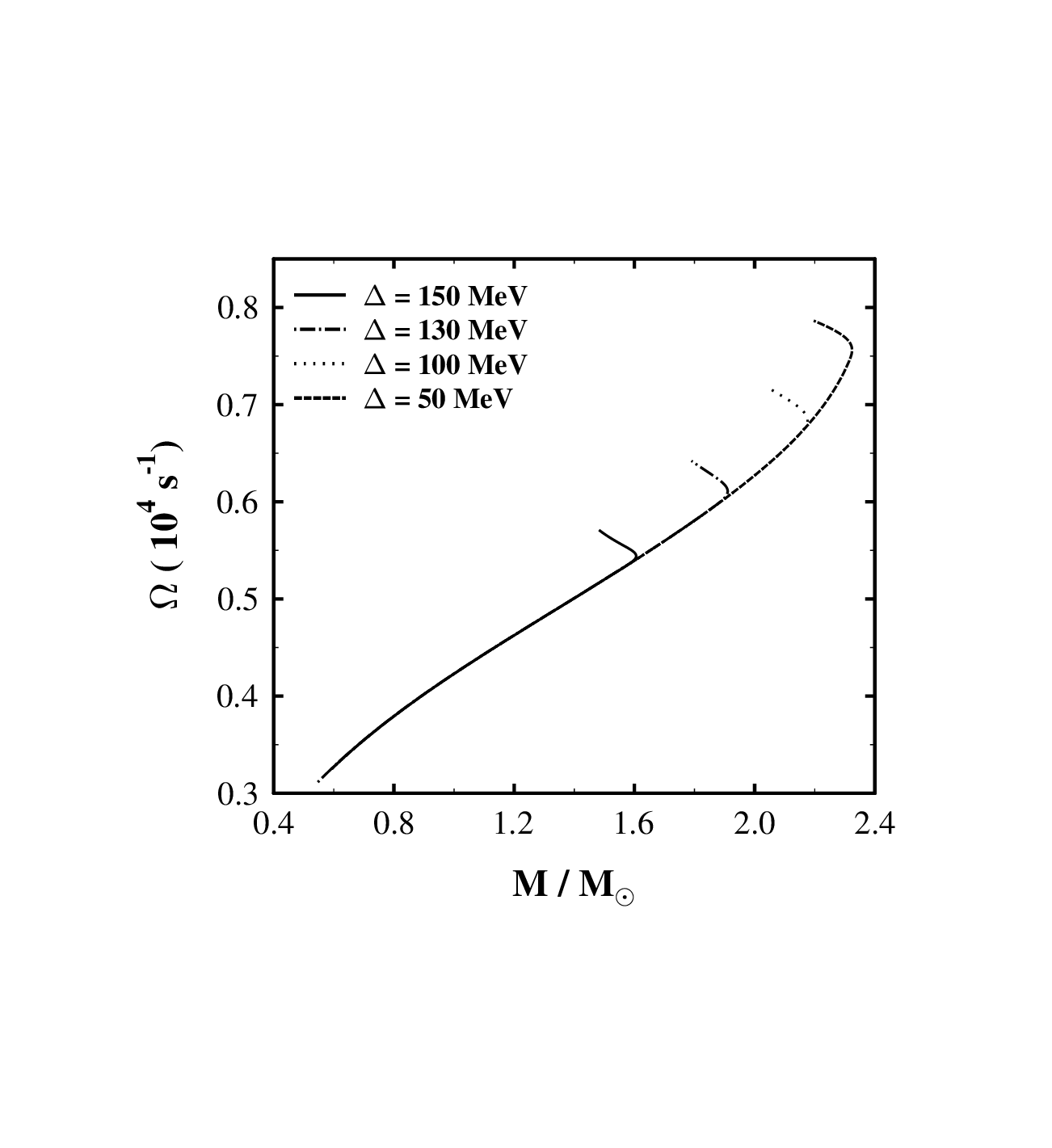}
\caption{Angular velocity versus the mass of the rotating compact
star for different values of pairing gap and for a fixed bag
pressure, $B^{1/4} = 210$ MeV.} \label{omegam}
\end{figure}
\begin{figure}
\includegraphics[width=10.cm]{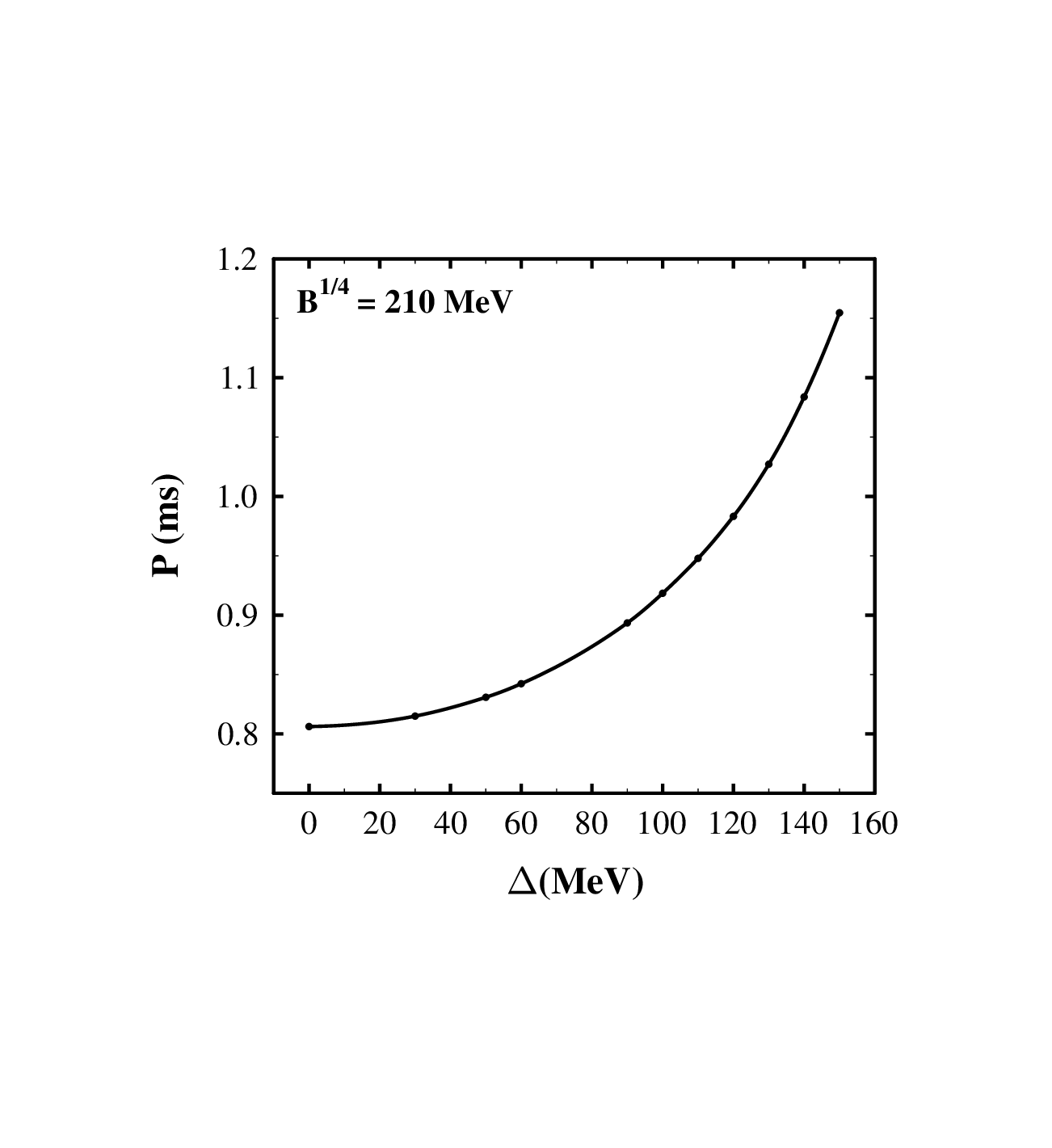}
\caption{The period of the rotating compact star as a function of
the pairing gap for a fixed bag pressure, $B^{1/4}= 210$ MeV. The
stars with low pairing gap seem to have sub-millisecond periods.}
\label{tdelta}
\end{figure}
\end{document}